# THE STRUCTURE OF THE CLIMATE DEBATE

*Richard S.J. Tol*

*Department of Economics, University of Sussex, Falmer, United Kingdom*

*Institute for Environmental Studies, Vrije Universiteit, Amsterdam, The Netherlands*

*Department of Spatial Economics, Vrije Universiteit, Amsterdam, The Netherlands*

*Tinbergen Institute, Amsterdam, The Netherlands*

*CESifo, Munich, Germany*

**Abstract**

First-best climate policy is a uniform carbon tax which gradually rises over time. Civil servants have complicated climate policy to expand bureaucracies, politicians to create rents. Environmentalists have exaggerated climate change to gain influence, other activists have joined the climate bandwagon. Opponents to climate policy have attacked the weaknesses in climate research. The climate debate is convoluted and polarized as a result, and climate policy complex. Climate policy should become easier and more rational as the Paris Agreement has shifted climate policy back towards national governments. Changing political priorities, austerity, and a maturing bureaucracy should lead to a more constructive climate debate.

## 1. Introduction

The best course of action for greenhouse gas emission reduction and adaptation to climate change has been subject to debate for three decades. Much less attention has been paid to the nature and structure of the climate debate, and why certain actors adopt the position that they do. In this paper, I use basic tools of political economy and the economics of organizations, as well as the philosophy of science and social psychology, to analyse the climate debate.

Economists have contributed large volumes of research on international environmental agreements and the architecture of international climate policy, and copious amounts of papers on the design of greenhouse gas emission reduction policies and policy instrument choice in the first- and second-best. Economists have been reluctant, however, to write much about the climate debate itself and apply their tools of analysis to the question why participants in this debate behave the way they do. This paper makes a first attempt.

Climate policy has moved slowly. This frustrates many, whether on the side of those who advocate for rapid emission reduction or of the opinion that climate policy is a nonsense best forgotten. In either case, it helps to understand why the discussion is as it is, why people argue as they do, and why climate policy has neither gone away nor moved to reduce greenhouse gas emissions. This paper makes a start in answering these questions. The method chosen is discursive rather than analytical, but readers who prefer analytics and econometrics just have to follow the references.

The paper proceeds as follows. Section 2 revisits the case for climate policy. Section 3 sketches the optimal design of climate policy. These sections are kept short because the material is well-rehearsed. Section 4 explores the positions taken by different actors in the climate debate. Section 5 discusses recent developments in international climate policy, with particular regard to the Paris Agreement and the replacement of targets and timetables by pledge and review. Section 6 treats national climate debates, in a necessarily cursory way as there are so many. Section 7 concludes.

## 2. The case for climate policy

Some people have argued that climate change is bad as all change is apparently for the worse (Potsdam Institute for Climate Impact Research and Climate Analytics 2014, WBGU 1995). This is an odd position: Nothing leads to good that is not natural. Gender equality would be a radical departure from the past, as would democracy in China, and universal access to sanitation. These decidedly unnatural things would generally be welcomed. Indeed, Hume (1740) has warned against using the world-as-is as a justification for how the world ought to be, and Moore (1903) against assuming that what-is-natural is good. However, as argued by Knappenberger and Michaels (2015), environmentalists typically claim that climate change is bad for all that is good and good for all that is bad.

The environmental movement's hankering for the presumed purity of times long gone and its resistance to progress reflects its roots in Romanticism (Hinchman and Hinchman 2007) as does its embrace of the naturalistic fallacy of von Schiller (1787) and the implied rejection of Hume (Miller 2005).

Environmentalists' construction of all climate change as necessarily bad is also at odds with its frequent embrace of scientific results as a key justification for environmental policy. Research has shown that climate change would bring both positive and negative impacts (Arent et al. 2014, Schneider et al. 2007, Smith et al. 2001, Pearce et al. 1996). Positive impacts include a reduced demand for energy for winter heating, fewer cold-related deaths, and carbon dioxide fertilization which makes crops grow faster and reduces their demand for water. Negative impacts include sea level rise, the spread of tropical diseases, and increases in storm intensity, droughts, and floods. Adding up all these impacts after having expressed them in welfare equivalents, the impact of initial climate change is probably slightly positive. This is irrelevant for policy, because initial climate change cannot be avoided. More pronounced climate change would have net negative effects, and these impacts would accelerate with further warming. Even so, the impacts would be moderate: The welfare impact of a century of climate change is comparable with the welfare impact of a year of economic growth (Tol 2015). Uncertainties are large, though, but even the most pessimistic estimates show that welfare loss due to a century of climate change is comparable to that of losing a decade of growth (Stern et al. 2006).

With relatively small numbers for the total impact of climate change, it is no surprise that a recent meta-analysis of estimates of the social cost of carbon concludes that its expected value is $201/tCO_2$ for a 0% pure rate of time preference, $^{2015}107/tCO_2$ for a 1% PRTP, and $13/tCO_2$ for 3% (Tol 2015). This compares to a price of carbon dioxide emission permits of $13/tCO_2$ in California[1] and of $5/tCO_2$ in the EU[2] in July 2016. These latter prices do not buy a lot of emission abatement.

## 3. The design of climate policy

Greenhouse gas emissions can be reduced in a number of ways (Clarke et al. 2014). More efficient energy use and a switch to alternative energy sources are the two main technical options, although reduced population growth, slower economic growth, carbon capture and storage, and geoengineering should be considered too.

Energy-saving and –switching are best stimulated by a carbon tax. Incentive-based policy instruments are better suited for reducing emissions from diffuse and heterogeneous sources than rule-based instruments (Baumol and Oates 1971). Taxes are more appropriate for stock pollutants than tradable permits (Weitzman 1974). A carbon tax is therefore the cheapest way to reduce greenhouse gas emissions. The initial carbon tax should be modest because it imposes a deadweight loss as it penalizes decisions made before there was a carbon tax (Wigley, Richels, and Edmonds 1996, Goulder and Mathai 2000). Over time, the carbon tax should rise. Net present abatement costs are lowest if all emissions from all sectors and all countries are taxed equally and if the carbon tax rises with a modified Hotelling rate (Hotelling 1931, Van Der Ploeg and Withagen 2014).

Higher carbon taxes would lead to deeper emission cuts. Only a modest carbon tax is needed to keep atmospheric concentrations below a high target but the required tax rapidly increases with the stringency of the target. If concentrations are to be kept below 450 ppm $CO_2$eq, the global carbon tax should reach some $210/tCO_2$ in 2020 or so (Tol 2013) – fifty times the recent price of permits in the Emissions Trading System which covers about half of emissions in Europe. Such a carbon tax would roughly double the price of energy in Europe. A 450 ppm $CO_2$eq concentration would give a 50/50 chance of meeting the declared goal of the European Union and the United Nations to keep global warming below 2°C (Peters et al. 2015, Peters 2016).

However, less ambitious targets would require far lower carbon taxes, and would hardly affect economic growth (Clarke et al. 2014, Clarke et al. 2009, Tavoni and Tol 2010). The above discussion about the impacts of climate change suggests that a modest carbon tax can be justified, but that more ambitious goals may be hard to defend (Tol 2013).

## 4. The debate on climate policy

I argue above that climate change is a relatively small problem that can easily be solved: We just need a modest carbon tax. A casual observer of climate policy and the media would have a different impression. Climate change is often presented in catastrophic terms (Hulme 2008), although there are also voices that decry it as a hoax (Jang and Hart 2015). Climate policy is often presented as costless if not beneficial (Barker et al. 2007), although there are also voices that claim it may bring the economy crashing down (Kelly 2016). These positions are untenable, so their persistence requires explanation.

---

[1] http://calcarbondash.org/
[2] https://www.eex.com/en#/en

A number of things stand in the way of a reasonable debate on international climate policy and the simple solution sketched above.

First, the presentation of climate change is often a discourse of fear (Hulme 2008). There is a demand for an explanation of the world in terms of Sin and a Final Reckoning This is often referred to as Millenarianism (Landes 2011). Although many Europeans are nominally secular, fewer are in practice. The story of climate change is often a religious one (Bruckner 2014): emissions (sin) lead to climate change (eternal doom); we must reduce our emissions (atone for our sins). This has led to an environmental movement (a priesthood) that thrives on preaching climate alarmism, often separated from its factual basis. Environmentalism further offers an identity (McCarthy 2002, Heise 2008), a tribe to belong to, and an opportunity to feel better than outsiders. In order to maximize their membership and income, environmental NGOs meet the demand for scaremongering and moral superiority (Bell 2015).

Second, climate policy is perfect for politicians. Climate change is a problem that spans the world and lasts for centuries. Substantial emission reduction requires decades and a degree of global cooperation. This offers the opportunity for politicians to channel their inner Bruce Willis and make grand promises about saving the world. At the same time, most of the burden of actually doing something – and hurting constituents – can be shifted to a successor. There are convenient foreigners to blame for current inaction. Climate policy is an ideal case to overpromise and underdeliver.

Climate change also provides opportunities to deflect attention. The Managing Director of the International Monetary Fund, Christine Lagarde, for instance, could be expected to take a keen interest in what is going on in the Eurozone and China – but she is remarkably eager to talk about climate change (Lagarde 2015, Rowley 2013, Lagarde 2013), an issue firmly outside the mandate of the IMF. As another example, the United State Secretary of State, John Kenny, regularly mentions climate change as his greatest worry (Goodell 2015, Almasy 2014, Kerry 2009) – in the USA, the Environmental Protection Agency takes the lead on climate policy – although others may argue that Islamic State, Russia and China are more pressing problems, and within the actual remit of the State Department.

Bureaucrats are a third reason why the climate debate is so convoluted. Emissions are best reduced through a carbon tax. A carbon tax is an excise duty. The legal, regulatory, administrative and logistical apparatus to levy excise duties is already in place, both in the governments that set the levies and in the companies that administer them. A carbon tax can thus be implemented with minimal effort. Bureaucrats, however, like to create new bureaucracies and expand existing ones (Niskanen 1971) – and climate policy has provided an opportunity to do exactly that. Climate policy has been a political priority for about two decades. Emissions have hardly budged, but a growing and by now vast number of civil servants have occupied themselves with creating a bureaucratic fiction that something is happening.

The international climate negotiations are an illustration. When the United Nations Framework Convention on Climate Change[3] (UNFCCC) first entered into force, two or three meetings a year were organized. Now, there are two or three meetings a week under the auspices of the UNFCCC.[4] Some 800 people attended the first Conference of the Parties in Berlin in 1995. There were 40,000 delegates in Copenhagen 2009. Twenty years of negotiations have not lead to a discernible reduction in emissions, but it has created a substantial body of obligations on reporting – which of course requires more staff in capital cities – while the multifaceted negotiations in multiple arenas requires careful preparation before the negotiations and evaluation after – again requiring more staff.

---

[3] See http://unfccc.int/files/essential_background/background_publications_htmlpdf/application/pdf/conveng.pdf

[4] See http://unfccc.int/2860.php

Bureaucrats are not alone in their desire for complications in climate policy. A uniform carbon tax is cost-effective as it minimizes friction in the economy. Market distortions imply rents (Krueger 1974, Bhagwati 1982). Some distortions are there to create rents for political allies, other rents may instil a sense of gratitude towards the politician who distorted the market. Politicians can also use climate policy to give subsidies or grant tax breaks that support certain households or companies, to impose technical that favour particular companies, and to steer grandparented emission permits preferentially towards specific emitters. Complications in climate policy thus serve the interests of rent seekers as well as the interests of policy makers who use rent creation to reward allies. Assessments of current and past climate policies indeed find that greenhouse gas emission reduction is not the main impact (Murray et al. 2014, Leahy and Tol 2012, Pearce 2006).

A fifth reason why climate policy is more complicated than needed is that it has been used to promote other agendas. Greenhouse gas emission reduction is a global public good (Hoel 1991, Carraro and Siniscalco 1992, Barrett 1990, van der Ploeg and de Zeeuw 1992, Battaglini and Harstad 2015, Stavins 2011). The costs of emission abatement are borne by the country that reduces the emissions. The benefits of emission reduction are shared by all of humankind. It is thus individually rational to do very little, and hope that others will do a lot. As every country reasons the same way, nothing much happens. There is no solution to this short of installing a world government (Biermann et al. 2012). Therefore, global climate policy has been used as a tactical argument by those who desire a world government for other reasons (Rockström et al. 2009). Similarly, climate policy has been used to transfer powers from Member States to the European Union: Environmental regulation, which is decided by qualified majority, shapes energy, transport and industrial policies, which are decided unanimously.

Because climate change is such a prominent issue, champions of other worthy causes too have joined the bandwagon. Climate policy is about gender (McCright and Dunlap 2011a, Kronsell, Smidfelt Rosqvist, and Winslott Hiselius 2016), about development (Olsen 2007), about fair trade (Viguier 2001, Köhler 2014, Ciscell 2010), about labour rights and employment (Yi 2013, Nugent 2011), and so on and so forth. The ultimate goal of climate policy – decarbonisation of the economy – is thus obscured. If climate policy serves multiple agendas, then the optimal policy no longer equates the marginal abatement costs (Tinbergen 1952). Climate policy should be reconfigured to meet different targets.

The final issue complicating climate policy is the following. Some object to a transfer of power to the United Nations or the European Union. Others are against more government regulation. People may not support the causes that have jumped on the climate bandwagon. And, of course, rents for some imply losses for others. Those people may of course propose a climate policy that is simpler and has an exclusive focus on greenhouse gas emissions. Alternatively, climate policy itself may be attacked.

Environmentalists and politicians have the unfortunate habit of phrasing the need for climate policy as a scientific imperative (Grundmann 2016, Grundmann 2007, Matthews 2016). The science has spoken, and we must act. This is a categorical error. Science tells you what would happen if. It does not tell us what to do (Pielke 2007). For instance, science tells us that water rapidly withdrawing from the beach is a sign of an impending tsunami. Science tells us that a run for the hills increases the chance of survival. Science does not tell us to run, our survival instinct does. Science tells us that the will to live is strong in most people, and a decision analysis would recommend to start running. But the imperative to run comes from our values rather than from our knowledge.

Similarly, but much more complicatedly, the imperative for climate policy is derived from our values rather than from our understanding of climate change and its impacts. Climate research has shown that unmitigated climate change has substantial consequences (Arent et al. 2014). Cost-benefit analyses have shown that these impacts justify greenhouse gas emission reduction (Nordhaus 1991). But the decision to do so remains a political one.

Whenever a politician says that the science commands action, the opposition could of course point out that this is a categorical error, and that alternative interpretations of the available evidence or other parameterizations of the felicity calculus would support a different course of action (Hulme 2009). That is a difficult argument to make – particularly in 30 seconds on prime time television or in 140 characters on Twitter. It is much easier to pick holes in the science-that-commands.

Climate research is a young discipline studying a complex subject, so there are genuine concerns about both the state of knowledge and the standard of evidence (Spencer 2008, Krauss and Von Storch 2012, Pielke 2013, Curry and Webster 2013, Lindzen 2012, 1990). At the same time, the desire to block climate policy is a clear incentive to exaggerate the things that we do not know and understand about climate change and its impacts, and to emphasize the unscholarly behaviour of a handful of climate researchers.

Some environmentalists have responded in kind to these challenges (Lewandowsky, Oreskes, et al. 2015, Lewandowsky, Oberauer, and Gignac 2013, Lewandowsky, Gignac, and Oberauer 2015, 2013, Lewandowsky, Cook, et al. 2015, Sills 2010, Gleick 2012, Oreskes et al. 2015, Mann 2015, Hansen 2013, Hauschild 2011). This has enabled further critique of climate research (Pielke 2010).

Both the climate debate (Hoffman 2011, Fisher, Waggle, and Leifeld 2013) and public opinion about climate change and policy (McCright and Dunlap 2011b, Kahan et al. 2012) are now strongly polarized. There are two implications of this. Small but vocal minorities can block new policy initiatives in countries where climate policy has not gotten off the ground. Australia and the USA are perhaps the key examples. In countries with climate policy, the European Union for example, the fractious debate on climate science crowds out scrutiny of climate policy as discussion on the merits of particular policy intervention often descend into mud-slinging about hiding the decline.

## 5. Progress in international climate policy

How to move forward? The Paris Agreement[5] of December 2015 is a major step in the right direction. Any solution to the climate problem should start with acknowledging that we live in a world of many countries, the majority of which jealously guards their sovereignty. Unfortunately, for 20 years, international climate negotiations under UNFCCC focussed on one thing: A legally binding treaty on targets for greenhouse gas emission reduction. The 1997 Kyoto Protocol[6] seemed to meet that aim, but although it did impose targets on most rich countries for a short period, sanctions were never agreed on countries that failed to reach their obligations. Key countries did not sign up anyway, while other countries met the agreed targets without much effort (Aichele and Felbermayr 2013, Bel and Joseph 2015, Löschel, Pothen, and Schymura 2015).

---
[5] See http://unfccc.int/files/essential_background/convention/application/pdf/english_paris_agreement.pdf
[6] See http://unfccc.int/resource/docs/convkp/kpeng.pdf

The Paris Agreement discarded legally binding emission targets. Instead, it is centred on Intended Nationally Determined Contributions to climate policy (INDCs). Article 4.2 has that "[e]ach Party shall prepare, communicate and maintain successive nationally determined contributions that it intends to achieve. Parties shall pursue domestic mitigation measures, with the aim of achieving the objectives of such contributions." In other words, the Paris Agreement obliges countries to have a climate policy – but the word "intended" signifies that climate policy is aspirational, while the words "nationally determined" specify that these aspirations are set by individual countries rather than through international negotiations.

Article 4.3 specifies that "[e]ach Party's successive nationally determined contribution will represent a progression beyond the Party's then current nationally determined contribution and reflect its highest possible ambition". Put differently, national aspirations should become more ambitious over time – but in an unspecified way and at an unspecified rate.

There are no sanctions for missing targets – indeed Article 15 explicitly states that the "mechanism to […] promote compliance" is "non-punitive". Article 28 stipulates that a country may withdraw from the Paris Agreement without sanction.

In other words, after 20 years of trying and failing to negotiate a treaty with legally binding emission reduction targets, the Paris Agreement switched to pledge and review. This is a welcome development. International law is weak, and cannot provide public goods. It is in each sovereign country's self-interest to free ride on the emission reduction efforts of other countries (Hoel 1991, Carraro and Siniscalco 1992, Barrett 1990, van der Ploeg and de Zeeuw 1992, Nordhaus and Yang 1996). Pledge and review is how other global public goods are provided (Bradford 2008). If a peace-keeping force is needed, or an emergency response to an epidemic, countries come together to assess the situation and what resources are deemed necessary. Each country then pledges whatever can be made available. The response is later reviewed and scaled up or down as needed. This is a messy and often inadequate way of doing things, but it sort-of-works – unlike the first 20 years of international climate policy which saw a steady rise of greenhouse gas emissions.

Pledge-and-review could imply unilateral climate policy, which is expensive. If a country raises its price of energy, but its trading partners do not, business will shift abroad. A country will be more ambitious if it is confident that its neighbours will adopt roughly the same climate policy. The annual meetings under the UNFCCC, and its internationally agreed standards on emissions monitoring and reporting take away most uncertainty about other countries' plans. If a carbon tax is the policy of choice, there would be a degree of implicit tax harmonization with key import and export markets (Baldwin and Krugman 2004, Mendoza and Tesar 2005) – perhaps further helped by informal discussions in regional trade organizations, such as the EU, NAFTA, MERCUSOR and ASEAN.

**6. Progress in national climate policies?**

The Paris Agreement puts climate policy firmly back in the hands of national governments. In principle, this should take away the concerns of those who see climate policy as a transfer of power to multilateral organizations. However, at least in Europe, politicians and environmentalists refer to a countries' international obligations under the Paris Agreement. There are none. The Paris Agreement only obliges countries to have a climate policy. It does not oblige them to have a stringent climate policy. But just like the presumed scientific imperative is a welcome excuse, so the alleged international obligation allows people to avoid making the argument for climate policy on its merits.

Politicians should not be afraid to argue for climate policy. Opinion polls in democratic countries have repeatedly shown over a period of 20 years that a majority is in favour of greenhouse gas emission reduction, even if that means more expensive energy (Leiserowitz et al. 2013, Leiserowitz, Kates, and Parris 2006, 2005, Lee et al. 2015). Cost-benefit analyses have consistently shown that a well-designed climate policy pays off (Nordhaus 1991, Nordhaus 1993, Tol 2013).

Changing the nature of national debates on climate policy is not straightforward. As shown above, all parties involved behave as they do for good reason. The general public, while still concerned about climate change, does not see it as a priority. As the ability of climate change to grab headlines dwindles, so does its appeal as a bandwagon to jump on. As climate bureaucracies have matured, so the opportunities for further rapid expansion have fallen. At the same time, political priorities have shifted – to immigration, Brexit, and fiscal and financial stability in Europe – and to race relations, immigration and the economy in the USA. That would mean fewer incursions by more senior but less knowledgeable politicians, so that climate policy would settle at a more technocratic, less politically volatile level of policy making.

Climate policy under the Obama administration is a good example (Burtraw 2015). The party of Nixon (who founded the Environmental Protection Agency), Reagan (who helped broker the Montreal Protocol on Substances that Deplete the Ozone Layer) and Bush the Elder (who helped negotiate the UNFCCC) is strongly against climate policy while the other side is in favour. The result is legislative deadlock. The EPA, however, has used – some say abused – the 1970 Clean Air Act (signed by Nixon) to begin to regulate greenhouse gas emissions. The EPA has used Clinton's Executive Order 12866 to ensure a uniform price of emission reduction equal to the social cost of carbon (Greenstone, Kopits, and Wolverton 2013, Pizer et al. 2014, Rose 2012, Sunstein 2014, Burtraw et al. 2014). Of course, the risk of policy by executive power is that the next president can readily reverse these steps, for instance by appointing different experts on the committee that sets the social cost of carbon (Dayaratna, McKitrick, and Kreutzer 2016, Havranek et al. 2015).

In Europe, the capital subsidies that are grandparented permits are gradually replaced by auctions, while austerity brought an end to overly generous, and unnecessary, subsidies on renewables in Southern Europe and the United Kingdom. But there is regress too. The UK's climate levy is on top of cap-and-trade; it raises costs but does not reduce emission (Boehringer, Koschel, and Moslener 2008). France recently introduced a similar tax. Germany has accelerated its nuclear phase-out, replacing it by coal and lignite. Resistance against natural gas from hydraulic fracturing remains strong.

Developing countries take their cues on climate policy from Europe and North America. There as well, patronage too often takes priority over emission reduction, and the various streams of climate funds are not without blemish (Mathy and Guivarch 2010, Michaelowa and Michaelowa 2007, Bhattacharyya, Intartaglia, and McKay 2016, Wara 2008, 2007). China is copying another European mistake. For stock problems like climatic change, taxation is superior to cap-and-trade (Weitzman 1974, Pizer 1997) but, after long deliberation, China opted for the latter (Hübler, Voigt, and Löschel 2014, Jotzo and Löschel 2014, Zhang et al. 2014, Karplus, Rausch, and Zhang 2016). There are two pre-conditions for successful permit trading. First, monitoring needs to be sound. We do not monitor carbon dioxide emissions. Instead, we impute carbon dioxide emissions from the volume and type of energy used. However, China's statistics on energy use are all over the place. Estimates of aggregate coal use, for instance, differ by twenty percent or more. Second, enforcement needs to be sound, also after an emission permit has changed hands a number of times. This is a concern in the EU-ETS (Aakre and Hovi 2010, D'Amato and Valentini 2011, Brandt and Svendsen 2014) which includes countries like Italy and Greece, but the legal system in China is weaker still.

In all this, it is important to give voice to those who see climate policy as a public policy that, like all public policies, can be designed well or not-so-well (McLure Jr 2014, Marron and Toder 2014, Goulder and Stavins 2011, Fischer and Fox 2011, Anthoff and Hahn 2010). Open and critical debate is key to a good design, as is independent evaluation of past policies. Delegitimizing dissenting voices by pejorative labelling and questioning motives is not part of an open discussion (Stern et al. 2016). Over the top rhetoric is not helpful (McKibben 2016).

## 7. Conclusion

First-best climate policy is simple: A uniform carbon tax, rising steadily over time, is all we need. Although an international agreement on a harmonized carbon tax is impossible, a degree of coordination through trade organizations and informal cooperation is feasible. A modest carbon tax can be justified based on conservative assumptions about people's preferences, and imposing such a carbon tax would avoid the worst of climate change without imposing much economic damage in the short run. In contrast, actual climate policy is not simple, and the debate on climate policy is polarized and convoluted. Environmental NGOs exaggerate climate change because that increases their membership and influence. Other advocates use climate as a bandwagon to pursue unrelated issues. Politicians use climate policy to create rents and hand-outs. For civil servants, climate policy is a way to expand bureaucracies. Those opposed to climate policy use weaknesses in climate science to derail the discussion. The Paris Agreement has shifted international climate policy away from legally binding targets to a more realistic system of pledge and review. This should take away some of the opposition to climate policy and allow policy makers to focus on national climate policies instead. Austerity, new political priorities, and a maturing bureaucracy should allow climate policy to move towards a focus on emission reduction.

This paper makes a number of observations about current and past climate policy that can and should be tested against data. Progress in natural language processing allow for hypotheses that were once in the realm of qualitative analysis to be tested rigorously and objectively. Less innovatively, despite a 25 year history of climate policy, taking the 1991 carbon tax in Norway as the starting point, there is dearth of ex post evaluations of what actually happened (Aghion et al. 2016, Commins et al. 2011). Such studies would show the costs of actual climate policies, rather than the costs of ideal climate policy, and reveal whether climate policy has objectives besides the stated one of greenhouse gas emission reduction. This paper also makes a number of predictions. Time will tell. The other points raised should be high on any research agenda on climate policy.

**Acknowledgements**